\pdfoutput=1

\documentclass[traditabstract]{aa} 

\usepackage{graphicx}

\usepackage{txfonts}
\usepackage{natbib}
\bibpunct{(}{)}{;}{a}{}{,} 
\usepackage{enumitem}

\begin{document}
   \title{Outflow Structure and Velocity Field of Orion Source I: ALMA Imaging of SiO Isotopologue Maser and Thermal Emission}

   \author{F. Niederhofer \and E.M.L. Humphreys \and C. Goddi}

   \institute{European Southern Observatory, Karl-Schwarzschild-Stra\ss e 2, D-85748 Garching bei M\"unchen, Germany\\
              \email{fniederh@eso.org, ehumphre@eso.org, cgoddi@eso.org}}

  \abstract
{Using Science Verification data from the Atacama Large Millimeter/Submillimeter Array (ALMA), we have identified and imaged five rotational transitions (J=5-4 and J=6-5) of the three silicon monoxide isotopologues $^{28}$SiO v=0, 1, 2 and $^{29}$SiO v=0 and $^{28}$Si$^{18}$O v=0 in the frequency range from 214 to 246 GHz towards the Orion BN/KL region. The emission of the ground-state $^{28}$SiO, $^{29}$SiO and $^{28}$Si$^{18}$O shows an extended bipolar shape in the northeast-southwest direction at the position of Radio Source~I, indicating that these isotopologues trace an outflow ($\sim$18 km~s$^{-1}$, P.A. $\sim$50$^{\rm o}$, $\sim$5000 AU in diameter) that is driven by this embedded high-mass young stellar object (YSO). Whereas on small scales (10-1000 AU) the outflow from Source I has a well-ordered spatial and velocity structure, as probed by Very Long Baseline Interferometry (VLBI) imaging of SiO masers, the large scales (500-5000 AU) probed by thermal SiO with ALMA reveal a complex structure and velocity field, most likely related to the effects of the environment of the BN/KL region on the outflow emanating from Source~I. 

The emission of the vibrationally-excited species peaks at the position of Source~I. This emission is compact and not resolved at an angular resolution of $\sim$1$\farcs$5 ($\sim$600 AU at a distance of 420 pc). 2-D Gaussian fitting to individual velocity channels locates emission peaks within radii of 100 AU, i.e. they trace the innermost part of the outflow. 
A narrow spectral profile and spatial distribution of the v=1 J=5-4 line similar to the masing v=1 J=1-0 transition, provide evidence for the most highly rotationally excited (frequency $>$ 200 GHz) SiO maser emission associated with Source~I known to date. The 
maser emission will enable studies of the Source~I disk-outflow interface with future ALMA longest baselines.}

\titlerunning{Outflow Structure and Velocity Field of Orion Source I}
   \maketitle
%
\section{Introduction}

Radio Source I in the Becklin-Neugebauer/Kleinmann-Low (BN/KL) region (d=418$\pm$6 pc; \citealt{Kim08}) provides the closest example of high-mass star formation. It is a deeply embedded young stellar object (YSO) and no infrared emission has been detected up to 22 $\mu$m \citep{Greenhill04a}. However, \citet{Okumura11} detected a mid-infrared counterpart to Source~I as a peak in the 7.8$\mu$m/12.4$\mu$m color temperature, indicating its protostellar nature. Measurements of the proper motions of Source I and the BN object, the brightest mid-infrared source with a bolometric luminosity of 1.3 $\times$ $10^4$L$_{\sun}$ \citep{deBuizer12} in this region, in combination with N-body simulations performed by \citet{Goddi11a}, suggest that Source I is a binary system with a total mass of $\sim$20 M$_{\sun}$ that sustained a close ($\sim$50 AU) encounter with the BN object about 500 years ago. This dynamical interaction caused  binary hardening (r $<$ 10 AU) while preserving the original circumbinary disk  around Source I  \citep{MoeckelGoddi12}, and ejection of the high-energy, high-velocity, wide-angle outflow in the Orion BN/KL region observed in molecular hydrogen and carbon monoxide \citep{Bally11,Zapata09}.
 
Source I powers a broad ensemble of SiO and H$_2$O masers within radii between 10 and 1000 AU and is one of only three star-forming regions known to host SiO masers \citep{Hasegawa86}. Vibrationally-excited v=1 and v=2 SiO masers trace the innermost part of Source I inside 100 AU. Very Long Baseline Interferometry (VLBI) observations revealed an X-shaped pattern of the SiO masers that brackets Source I \citep{Greenhill98,Kim08,Matthews10}. In this structure, the north and the west arms are red-shifted, whereas the south and the east arms are blue-shifted. Additionally \citet{Matthews10} found a bridge traced only by the v=2 masers that connects the south and the west arm and shows an intermediate velocity. The emission from the v=1 and v=2 SiO is interpreted as arising from the limbs of a wide angle, funnel-like, slow ($\sim$18 km~s$^{-1}$), bipolar outflow with a northeast-southwest axis that is launched from a rotating edge-on disk \citep{Greenhill04b,Matthews10}. \citet{Reid07} observed a $\lambda$7 mm elongated continuum emission of $\sim$50 AU in radius that originates from the center of the SiO maser emission, interpreted as an ionized disk elongated northwest-southeast. Monitoring of the $\lambda$7 mm continuum emission reveals a remarkably constant flux density and shape over a timescale of about 9 years, supporting the ionized disk model \citep{Goddi11a}.

Downstream the flow, at radii between 100-1000 AU, masers from H$_2$O and vibrational groundstate SiO as well as thermal SiO emission trace a bipolar collimated, low density outflow originating from Source I \citep{Greenhill98,Greenhill12,Plambeck09}. \citet{Greenhill12} related this large-scale outflow to a continuation of the wind emanating from the disk, making Source~I an excellent case of a disk-wind collimated in a bipolar outflow. VLBI monitoring of SiO masers enabled for the first time a spatially resolved study of the region where the wind is launched. Although we have a relatively good knowledge about this inner region, the large scale structure of the outflow remains ambiguous. On scales $>$ 1000 AU, the outflow shows a complex morphology and a velocity field that is inconsistent with the simple picture of a linear, collimated and rotating outflow. Observations of the SiO v=0 J=2-1 line performed with CARMA by \citet{Plambeck09} show increasing absolute line-of-sight velocities with distance away from Source~I and extended emission almost perpendicular to the outflow (cf. their Figure 3). \citet{Plambeck09} interpreted this emission as due to a precession of the outflow on timescales of a few hundred years. This timescale is, however, inconsistent with a hard binary with an orbital separation of a few AUs and an orbital period of a few years \citep{Goddi11a,Bally11}. \citet{Zapata12} reported a subset of the ALMA data shown here, for the J=5-4 v=0 SiO line only, suggesting rotation of the outflow. The proposed rotation, however, is in the opposite sense to the rotation measured on smaller scales with SiO J=1-0 masers and which therefore challenges their interpretation.

We report results from an ALMA
Science Verification (SV) spectral survey at 214-246 GHz towards the Orion BN/KL region. We identified and mapped 5 transitions of SiO isotopologues in the outflow of Source I. Our main goal was to study in detail the large-scale structure and velocity field of the Source~I outflow. These data also enabled us to find the first evidence for high-frequency ($>$ 200 GHz) SiO maser emission at the position of Source~I. 

The structure of this paper is the following. In Section \ref{sec:obs} the details of the observations as well as the imaging technique are described. The results are given in Section \ref{sec:res}. In Section \ref{sec:disc} we discuss the structure and the velocity field of the outflow of Source I as well as the nature of the SiO emission detected with ALMA. Conclusions and future work prospects are drawn in Section \ref{sec:concl}.

\section{Observations and Imaging\label{sec:obs}}

\begin{table*} \scriptsize
\caption{ALMA Observational Parameters\label{ALMA_Obs}}
\centering
\begin{tabular}{l c } 
\hline\hline
\noalign{\smallskip}
ALMA Observational Parameters & \\
\noalign{\smallskip}
\hline
\noalign{\smallskip}
Number of Antennas & 16 in compact ES Cycle 0 configuration \\
Frequency Range & 214 -246 GHz (lower 2/3 of total frequency range in ALMA Band 6)\\
Weather Conditions & PWV (mm) $\sim$2.3-2.8 (Obs. Setup 1), $\sim$1.8-2.6 (Obs. Setup 2), $\sim$1.8-2.5 (Obs. Setup 3)\\
\textbf{Synthesized Beam} & $\sim$1$\farcs$7$\times$1$\farcs$2\\
Number of Spectral Windows & 20 (3,840 channels each)\\
Spectral Resolution & $\sim$0.7 km~s$^{-1}$ \\
Theoretical Sensitivity &0.01 Jy beam$^{-1}$ in 0.7  km~s$^{-1}$\\
\noalign{\smallskip}
\hline
\end{tabular}
\end{table*}

The ALMA SV spectral survey of the Orion KL Hot Core Region was conducted on 2012 January 20 and covered the frequency range 214-246 GHz which corresponds to the lower 2/3 of the total frequency range in ALMA Band 6. 16 antennas (each 12 meters in diameter) were included in the compact Early Science (ES) Cycle 0 configuration. Three observational runs were performed consecutively. The precipitable water vapor (PWV) values during the individual runs were $\sim$2.3-2.8, $\sim$1.8-2.6 and $\sim$1.8-2.5 mm respectively. Each observation consists of five different spectral setups including 4 spectral windows each. The resulting 214-246 GHz frequency range is covered by 20 spectral windows of 1.875 GHz with an overlap of $\sim$0.2 GHz. Each spectral window consists of 3840 channels that are 488.281 kHz wide. This corresponds to a Hanning smoothed resolution in velocity of $\sim$0.7 km~s$^{-1}$.  
The theoretical line sensitivity of the observations, estimated with the ALMA Sensitivity Calculator, is 0.01 Jy~beam$^{-1}$ in 0.7 km~s$^{-1}$. Table \ref{ALMA_Obs} summarizes the observational parameters.

The ALMA SV data were calibrated prior to public release. 
Callisto served as the absolute flux calibrator and the quasar J0607-085, with a flux of 1.4 Jy, was used as both bandpass and phase calibrator. 
For imaging we used the Common Astronomy Software Applications (CASA). 
The first step of the SiO line imaging process was the continuum subtraction. We identified all channels with no line emission in each spectral window containing an SiO line and used the \textit{uvcontsub} 
task which fits a model to the continuum and subtracts it afterwards. For the cleaning of the line datasets we used a pixel size of 0$\farcs$3, an imagesize of 30$\arcsec$ $\times$ 30$\arcsec$ and a Briggs weighting with a robustness parameter of 1. This provided a synthesized beam of $\sim$1$\farcs$7 $\times$ 1$\farcs$2 at a P.A. of $\sim$1$^{\rm o}$. In each data cube, we used the strongest emission feature to perform a single step, phase-only self-calibration. Tests with further steps of self-calibration did not improve the image quality significantly. The rms noise in the line-free channels was about 0.01 Jy~beam$^{-1}$ while it was up to 0.3 Jy~beam$^{-1}$ in the channels with the strongest emission.

\section{Results\label{sec:res}}

\begin{figure}[h]
   \centering
   \includegraphics[width=8cm]{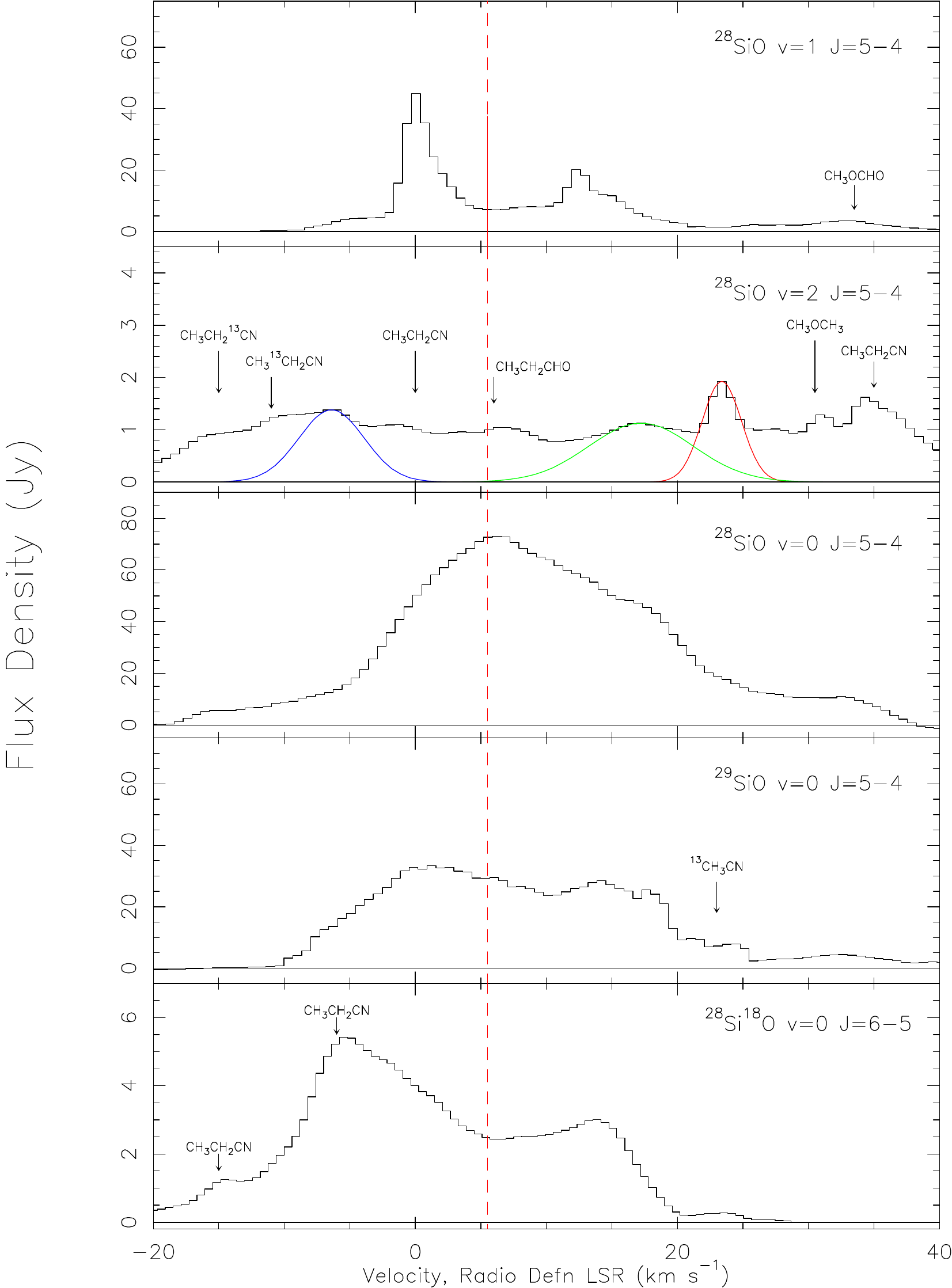}
   \caption{Spectra of the five  J=5-4 (SiO) and J=6-5 (Si$^{18}$O) transitions from SiO isotopologue  observed with ALMA. The dashed red line is at the systemic velocity of 5.5 km~s$^{-1}$. To guide the readers eyes, we overplotted the $^{28}$SiO v=2 spectrum with three Gaussians of arbitrary width to highlight the three peaks that are due to the SiO emission.}
              \label{ALMA_spectra}
    \end{figure}

\begin{figure*}
   \centering
   \includegraphics[width=18cm]{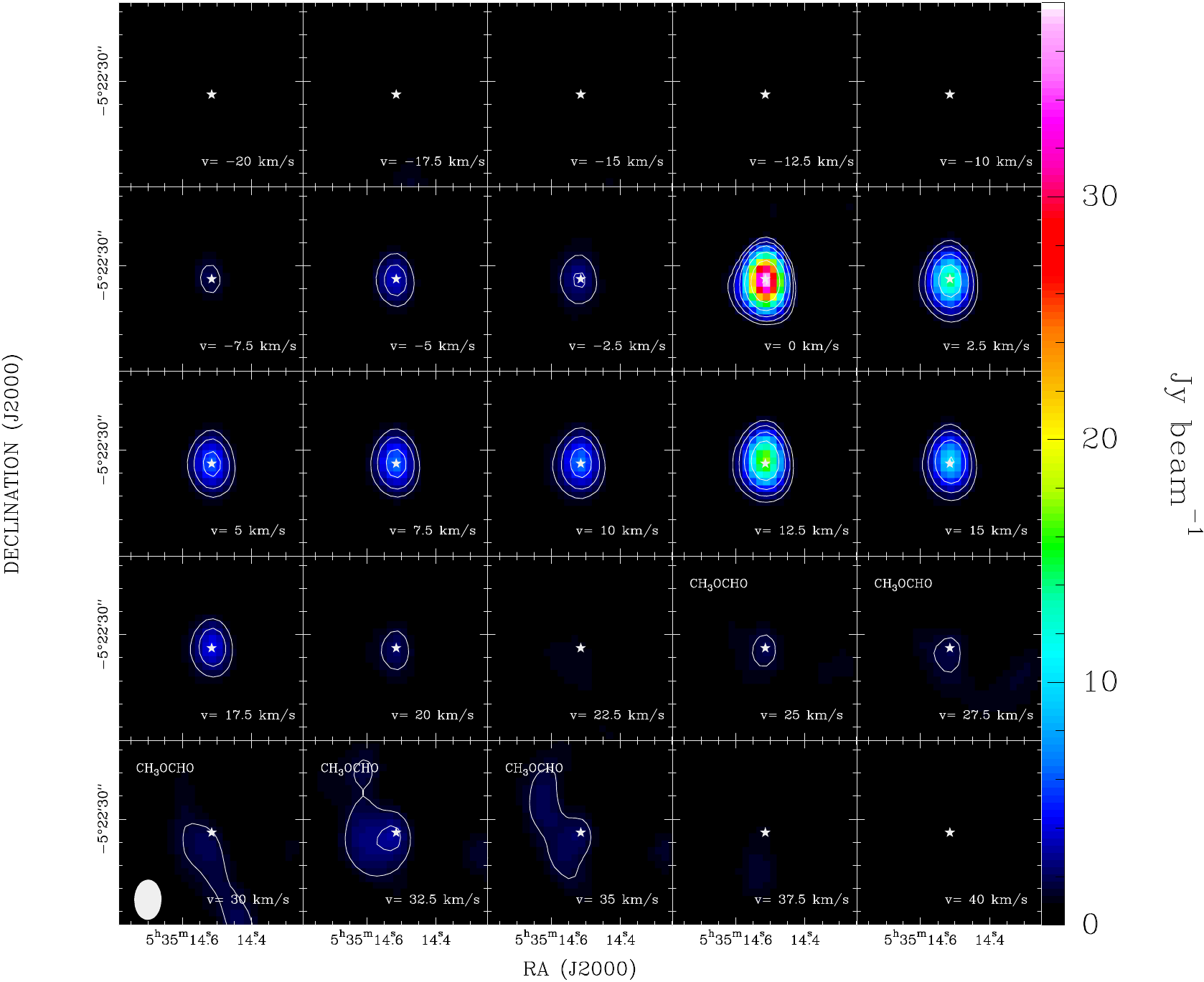}
      \caption{Channel maps of the $^{28}$SiO v=1 J=5-4 line emission. The white contours are at 4, 8, 16, 32, 64 and 128 times the rms noise in the peak line channel (0.29 Jy~beam$^{-1}$). Every channel is 2.5 km~s$^{-1}$ wide and the central LSR velocity is indicated in the lower right corner of each box. 
The two narrow emission features in the spectral profile at around 0 km~s$^{-1}$ and 12 km~s$^{-1}$ are clearly evident in the channel maps.
The emission from the v=1 $^{28}$SiO line in the central channel maps is unresolved and  centered at Source I (marked with a white star in each channel). 
The more diffuse emission at velocities $>25$ km~s$^{-1}$ instead does not peak at the position of Source~I and is likely associated with CH$_3$OCHO.  The ALMA synthesized beam is indicated as a white ellipse in the lower left panel.
              }
         \label{28SiOv=1_data_cube}
   \end{figure*}

\begin{figure*}
   \centering
   \includegraphics[width=18cm]{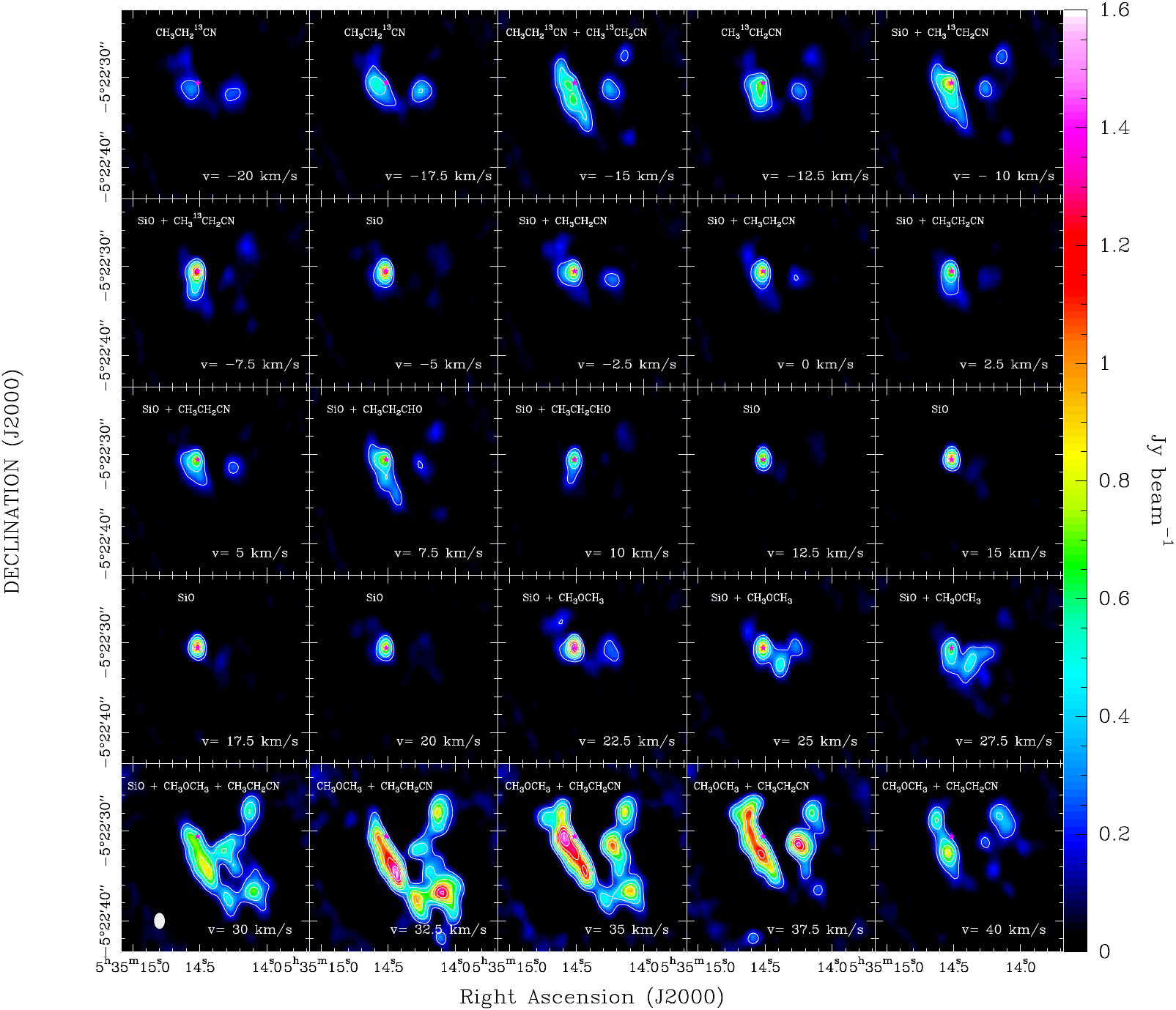}
      \caption{Channel maps of the $^{28}$SiO v=2 J=5-4 line emission. The contours in each channel are 3, 6, 9, 12, 15, 18, 21 and 24 times the rms noise in the peak line channel (0.065 Jy~beam$^{-1}$). Every channel is 2.5 km~s$^{-1}$ wide and the central LSR velocity is indicated in the lower right corner of each box. In the central channels, the SiO emission is unresolved and peaks at the position of Source~I (pink star symbol), probably tracing the inner parts of the outflow associated with Source I (as the $^{28}$SiO v=1 transition). In addition, strong diffuse emission is evident away from the systemic velocity blending the SiO line, that we attribute to CH$_3$CH$_2$CN, CH$_3$CH$_2$CHO and CH$_3$OCH$_3$. 
 The ALMA synthesized beam is indicated as a white ellipse in the lower left panel.
              }
         \label{28SiOv=2_data_cube}
   \end{figure*}

\begin{figure*}
   \centering
   \includegraphics[width=18cm]{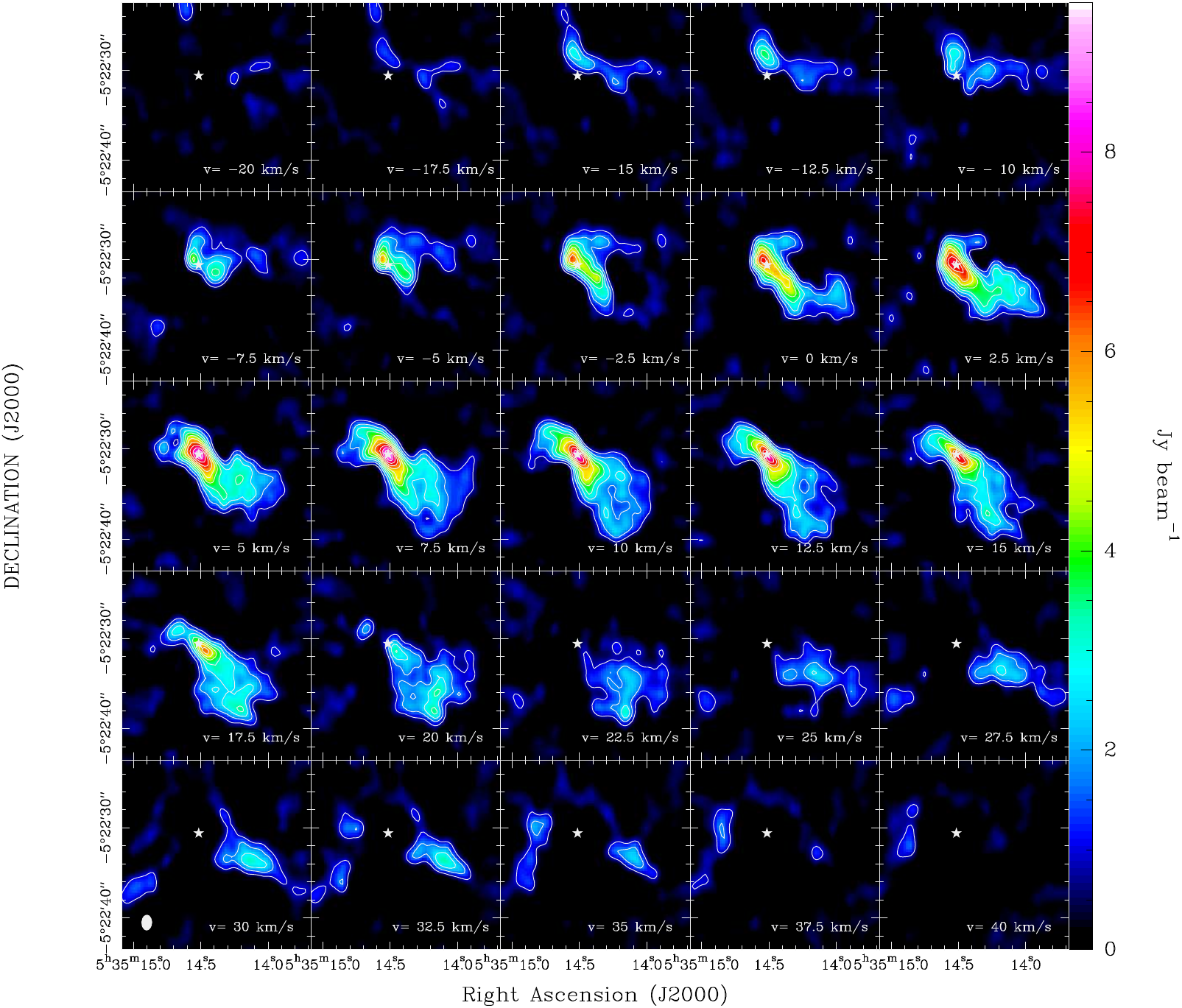}
      \caption{Channel maps of the $^{28}$SiO v=0 J=5-4 line emission. 
The contours in each channel are 3, 6, 9, 12, 15, 18, 21, 24, 27, 30 and 33 times the rms noise in the peak line channel (0.27 Jy~beam$^{-1}$). 
Every channel is 2.5 km~s$^{-1}$ wide and the central LSR velocity is indicated in the lower right corner of each box. 
The bulk of the emission  comes from an extended structure elongated  northeast-southwest and centered at Source~I (white star in each channel). The ALMA synthesized beam is indicated as a white ellipse in the lower left panel.
              }
         \label{28SiOv=0_data_cube}
   \end{figure*}

\begin{figure*}
   \centering
   \includegraphics[width=18cm]{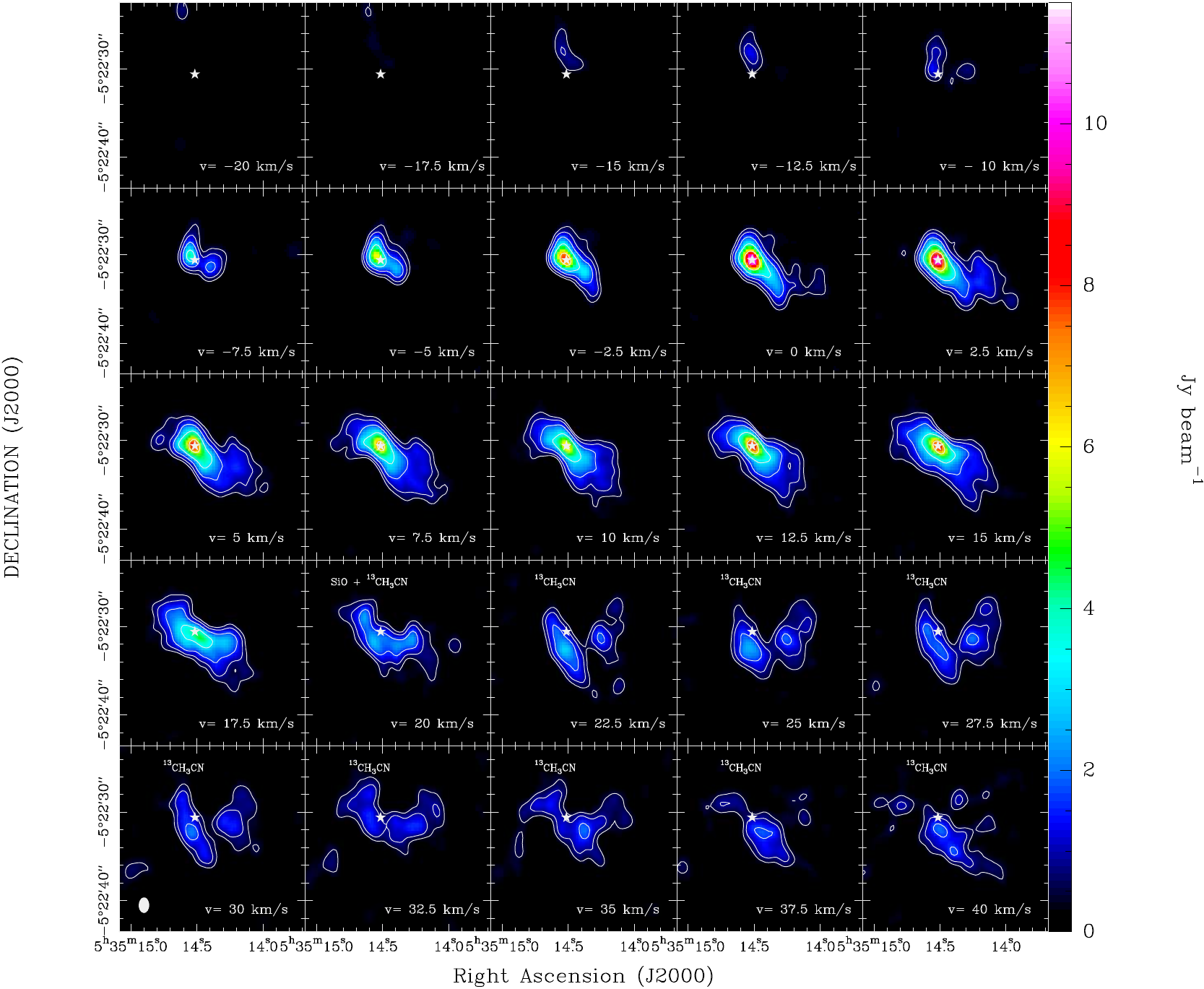}
      \caption{Channel maps of the $^{29}$SiO v=0 J=5-4 line emission. 
The contours in each channel are 4, 8, 16, 32 and 64 times the rms noise in the peak line channel (0.1 Jy~beam$^{-1}$). 
Every channel is 2.5 km~s$^{-1}$ wide and the central LSR velocity is indicated in the lower right corner of each box.
The emission peaks at Source I (white star) and shows a bipolar structure elongated northeast-southwest. 
It is double-peaked in velocity space with maxima at 1 km~s$^{-1}$ and 13 km~s$^{-1}$. 
The ALMA synthesized beam is indicated as a white ellipse in the lower left panel.
              }
         \label{29SiOv=0_data_cube}
   \end{figure*}

\begin{figure*}
   \centering
   \includegraphics[width=18cm]{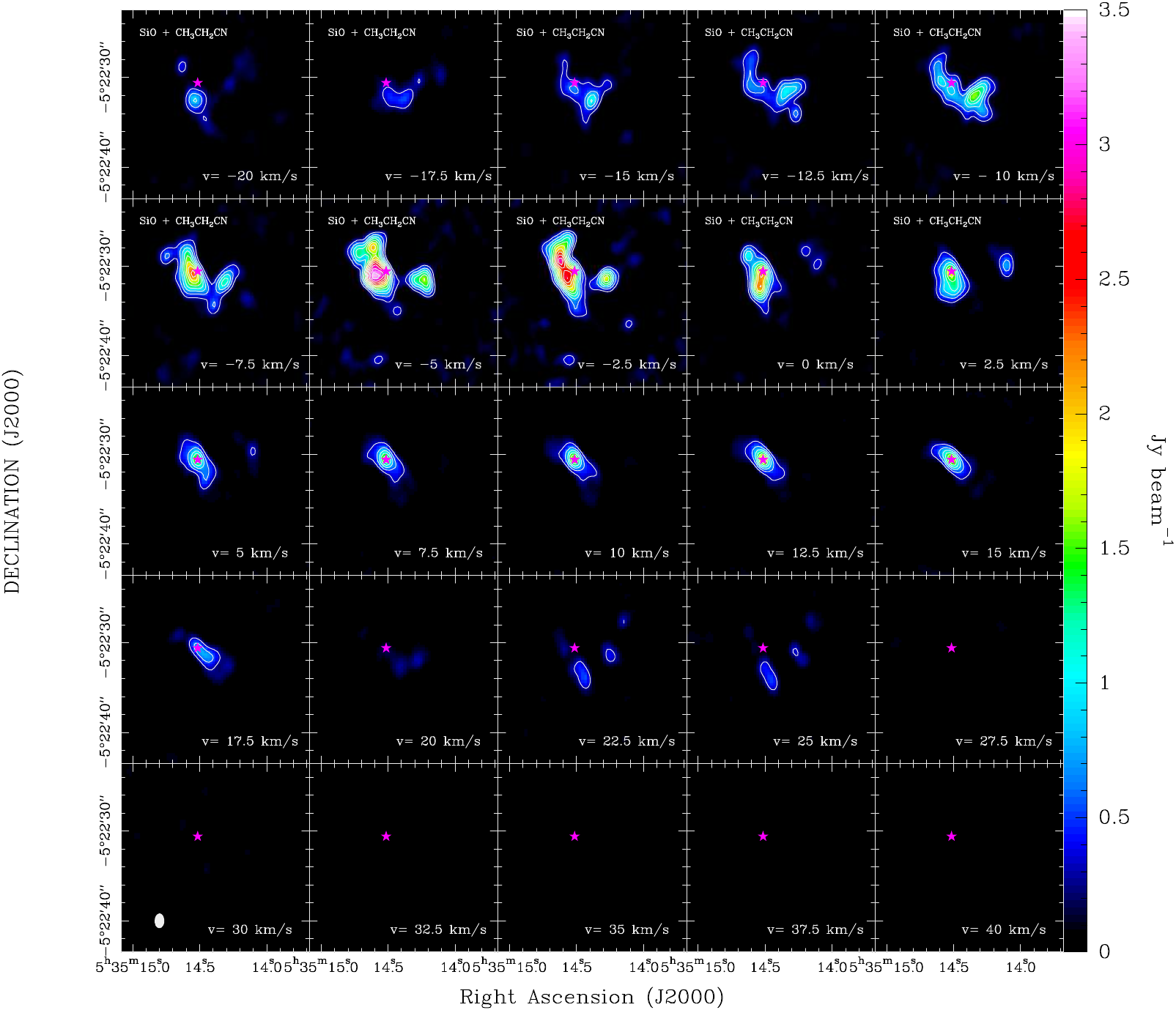}
      \caption{Channel maps of the $^{28}$Si$^{18}$O v=0 J=6-5 line emission. 
The contours are at 4, 8, 12, 16, 20, 24, 28, 32, 36, 40, 44 and 48 times the rms noise in the peak line channel (0.07 Jy~beam$^{-1}$). 
Every channel is 2.5 km~s$^{-1}$ wide and the central LSR velocity is indicated in the lower right corner of each box.
 The emission from the $^{28}$Si$^{18}$O isotopologue is mostly compact and peaks on Source I (pink star), but displays also a low-brightness elongated structure in the northeast-southwest direction, indicating that the $^{28}$Si$^{18}$O isotopologue traces the same outflow structures as the $^{28}$SiO v=0 and $^{29}$SiO v=0 lines. 
The diffuse and extended emission at larger velocities mostly originates from the hot core, south of Source I, and may be attributed to CH$_3$CH$_2$CN.
The ALMA synthesized beam is indicated as a white ellipse in the lower left panel. 
              }
         \label{28Si18Ov=0_data_cube}
   \end{figure*}

We detected and imaged five lines in total from the three SiO isotopologues $^{28}$SiO, $^{29}$SiO and $^{28}$Si$^{18}$O. All lines are J=5-4  rotational transitions   except for the J=6-5 transition of the $^{28}$Si$^{18}$O isotopologue. The $^{29}$SiO and $^{28}$Si$^{18}$O isotopologue transitions are in the vibrational ground-state whereas the $^{28}$SiO is detected in the ground-state as well as in the excited states v=1 and v=2.
The parameters of these data are given in Table \ref{ALMA_Data}. Figure \ref{ALMA_spectra} shows the spectral profiles of the individual transitions, which appear in some cases to be blended with other molecular transitions. Complementary spatial information provided by the interferometric images in different velocity channels helped us distinguish between different molecular lines and simplified the line identification. The ``contaminants'' are indicated in the spectrum with the name of the molecule at the respective line frequencies. The channel maps for each transition are given in the Figures \ref{28SiOv=1_data_cube} - \ref{28Si18Ov=0_data_cube}. The channels are 2.5 km~s$^{-1}$  wide (velocity-averaged for imaging purposes) and cover the velocity range from -20 km~s$^{-1}$ to 40 km~s$^{-1}$. 

\begin{table*} \scriptsize
\caption{Parameters of the Emission Lines of the isotopologues $^{28}$SiO, $^{29}$SiO and $^{28}$Si$^{18}$O observed with ALMA \label{ALMA_Data}} 
\centering
\begin{tabular}{l c c c c c c}
\hline\hline
\noalign{\smallskip}
Molecule & Restfrequency & Transition & Synthesized Beam & Peak flux & Line Width \tablefootmark{a}  & Brightness Temperature \\
 & (GHz) & J$_{u}$-J$_{l}$ & $\theta_M(\arcsec)\times\theta_m(\arcsec)$; P.A.($^{\rm o}$) & (Jy) & (km~s$^{-1}$) & (K) \\
\noalign{\smallskip}
\hline
\noalign{\smallskip}
$^{29}$SiO v=0     	&  214.38574	&  5-4	&	1.76$\times$1.18; 0.91	& 33.46	& 28	& $\geq$100\\
SiO v=2     		&   214.08854	&  5-4	& 	1.77$\times$1.19; 0.46	& 2.25	& 29\tablefootmark{b}	&...\\
SiO v=1    		&   215.59595	&  5-4	&	1.75$\times$1.18; 0.98	& 44.97	& 28	& $\geq$580\\
SiO v=0     		&  217.10498	&  5-4	&	1.72$\times$1.17; 1.29	& 72.96	& 55 	& $\geq$50\\
Si$^{18}$O v=0		& 242.09493     &  6-5	&	1.63$\times$1.08; 1.03	& 5.42	& 19\tablefootmark{b}	&...\\
\noalign{\smallskip}
\hline
\end{tabular}\\
\tablefoottext{a}{Quoted is full width at zero maximum} \\
\tablefoottext{b}{We estimated a lower limit of the velocity range, however due to line blending the line may in fact be broader}
\end{table*}

\subsection{The $^{28}$SiO v=1 Line}
The spectrum of the $^{28}$SiO v=1 J=5-4 line is clearly double peaked with the blue-shifted peak being stronger than the red-shifted one (Figure \ref{ALMA_spectra}). The velocity range of the emission is $\sim$-8 km~s$^{-1}$ to $\sim$+20 km~s$^{-1}$. The peaks themselves, located at 0 km~s$^{-1}$ and 12.2 km~s$^{-1}$ respectively, are quite narrow with FWHM values of $\sim$3.5 km~s$^{-1}$. The high velocity feature at the red side of the line emission is likely due to methyl-formate (CH$_3$OCHO). \citet{Goddi09a,Goddi09b} report observations of the J=1-0 line from SiO isotopologues towards Orion BN/KL at several epochs using the GBT and the VLA. For the masing $^{28}$SiO v=1 J=1-0 transition, they found a similar double peaked spectral profile at all epochs with peaks at around -5 km~s$^{-1}$ and 15 km~s$^{-1}$ (cf. Figure 1 in \citealt{Goddi09b}). The ALMA image shows that the v=1 J=5-4 emission is peaked at the position of Source I and is not resolved with a beam of 1$\farcs$7$\times$1$\farcs$2 ($\sim$700$\times$500 AU at 420 pc) at 215 GHz (Figure \ref{28SiOv=1_data_cube}).

\begin{figure}[h]
\centering
\includegraphics[width=9cm]{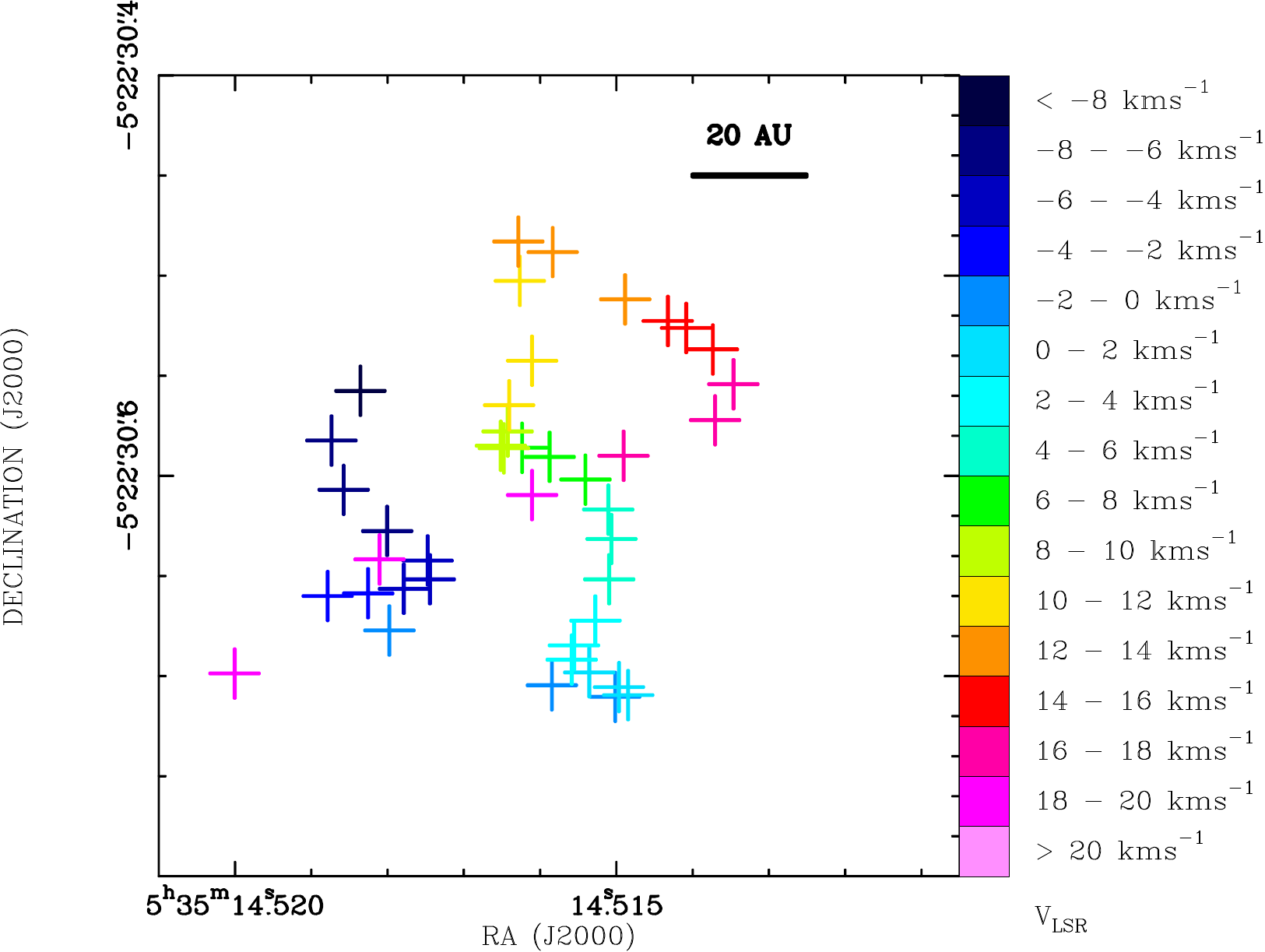}
\caption{Fitted positions of the $^{28}$SiO v=1 emission. The crosses are color coded by velocity. We see that there is a clear velocity gradient along the northwest-southeast direction, with blue-shifted emission in the eastern part and red-shifted emission in the western part. This is in good agreement with the result obtained by \citet{Matthews10} from VLBA data for the v=1 J=1-0 transition. The size of the image is about 160 AU x 160 AU. \label{spot_pos}}
\end{figure}

In order to characterize the structure and the kinematics of the $^{28}$SiO v=1 emission, we fitted a two-dimensional elliptical Gaussian model to the unresolved emission, in every velocity channel with emission greater that 4 times the rms noise in the peak line channel (i.e. 1.16 Jy~beam$^{-1}$), in at least six connected pixels. Since individual spots at similar velocities and different locations within the ALMA beam are blended together, the Gaussian fits give the intensity-weighted mean position of all the emission features in every velocity channel. Relative positional errors are given by $\delta \theta=0.5\theta_B/SNR$, where $\theta_B$ is the FWHM of the synthesized beam. Errors range from 10 mas for the brightest channels to 50 mas in the channels with weakest emission. The fitted positions are shown in Figure \ref{spot_pos}. We see a clear velocity gradient in the northwest-southeast direction, with the blue-shifted emission being in the east and the red-shifted in the west. The same blended structure is seen in the v=1,2 $^{28}$SiO and $^{29}$SiO and $^{30}$SiO v=0 J=1-0 transitions observed with the VLA by \citet{Goddi09b}. The velocity structure obtained from our Gaussian fits is consistent with the one found for the masing $^{28}$SiO v=1 and v=2 J=1-0 transitions by \citet{Matthews10} despite the fact that we cannot resolve the detailed four-armed structure of the wind with our ALMA synthesized beam of 1$\farcs$7$\times$1$\farcs$2. Nevertheless, the fitted positions are all located within $\sim$100 AU from Source~I. Hence, the vibrationally-excited v=1 J=5-4 transition traces the innermost part of the outflow associated with Source I, similarly to the v=1 and v=2 J=1-0 transitions \citep{Matthews10}.

\subsection{The $^{28}$SiO v=2 Line}

The spectrum associated with the $^{28}$SiO v=2 J=5-4 line is quite broad and shows several peaks (Figure \ref{ALMA_spectra}). However, only the three features at -6 km~s$^{-1}$, +17 km~s$^{-1}$ and +23 km~s$^{-1}$ are due to SiO emission. To guide the readers eyes, we have overplotted the spectrum with three Gaussians of arbitrary width at the positions of each SiO line peak. The channel maps of the emission are given in Figure \ref{28SiOv=2_data_cube}. The SiO v=2 emission is clearly unresolved, is peaked at the position of Source~I, and probably also traces the innermost part of the outflow, similar to the J=1-0 v=2 line \citep{Matthews10}. In some channels, we also see diffuse and quite strong  emission that we attribute to ethyl-cyanide (CH$_3$CH$_2$CN), dimethyl-ether (CH$_3$OCH$_3$) and propanal (CH$_3$CH$_2$CHO). This emission is blended with the v=2 SiO line and contributes to the remaining peaks in the spectrum. From the position of the peaks in the spectrum, we can set an interval from -6 km~s$^{-1}$ to +23 km~s$^{-1}$ as a lower limit to the velocity range of the $^{28}$SiO v=2 emission. The spectrum of the $^{28}$SiO v=2 J=1-0 transition obtained with the VLA by \citet{Goddi09b}, also shows a triple peaked  profile, however with peaks at approximately -6 km~s$^{-1}$, +6 km~s$^{-1}$ and +20 km~s$^{-1}$ (cf. their Figure 1). 

\subsection{The $^{28}$SiO v=0 Line \label{Res28SiOv=0}}

\begin{figure}
\centering
\includegraphics[width=9cm]{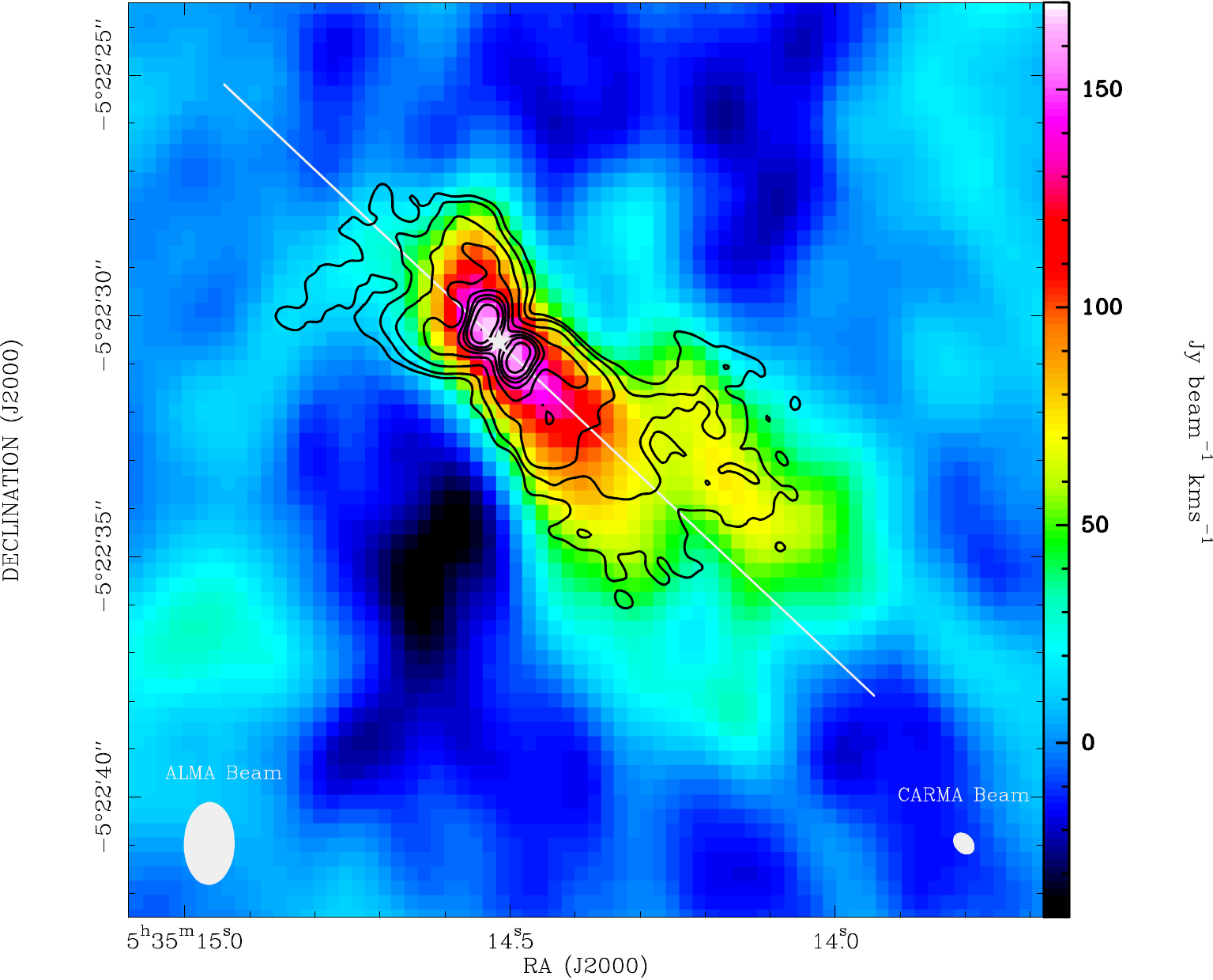}
\caption{Total intensity (moment 0) map of the v=0 J=5-4 $^{28}$SiO line. Overplotted are contours of the v=0 J=2-1 $^{28}$SiO total intensity from \citet{Plambeck09} obtained with CARMA. Contours are at 0.1, 0.2, 0.4, 0.7, 1.1, 1.5, 1.9, 2.6 and 3.7 Jy~beam$^{-1}$~km~s$^{-1}$. The white star marks the position of Source~I. The axis of the outflow (P.A. $\sim$50$^{\rm o}$) is indicated by a white line. \label{28SiOv=0mom0}}
\end{figure}

\begin{figure}
\centering
\includegraphics[width=9cm]{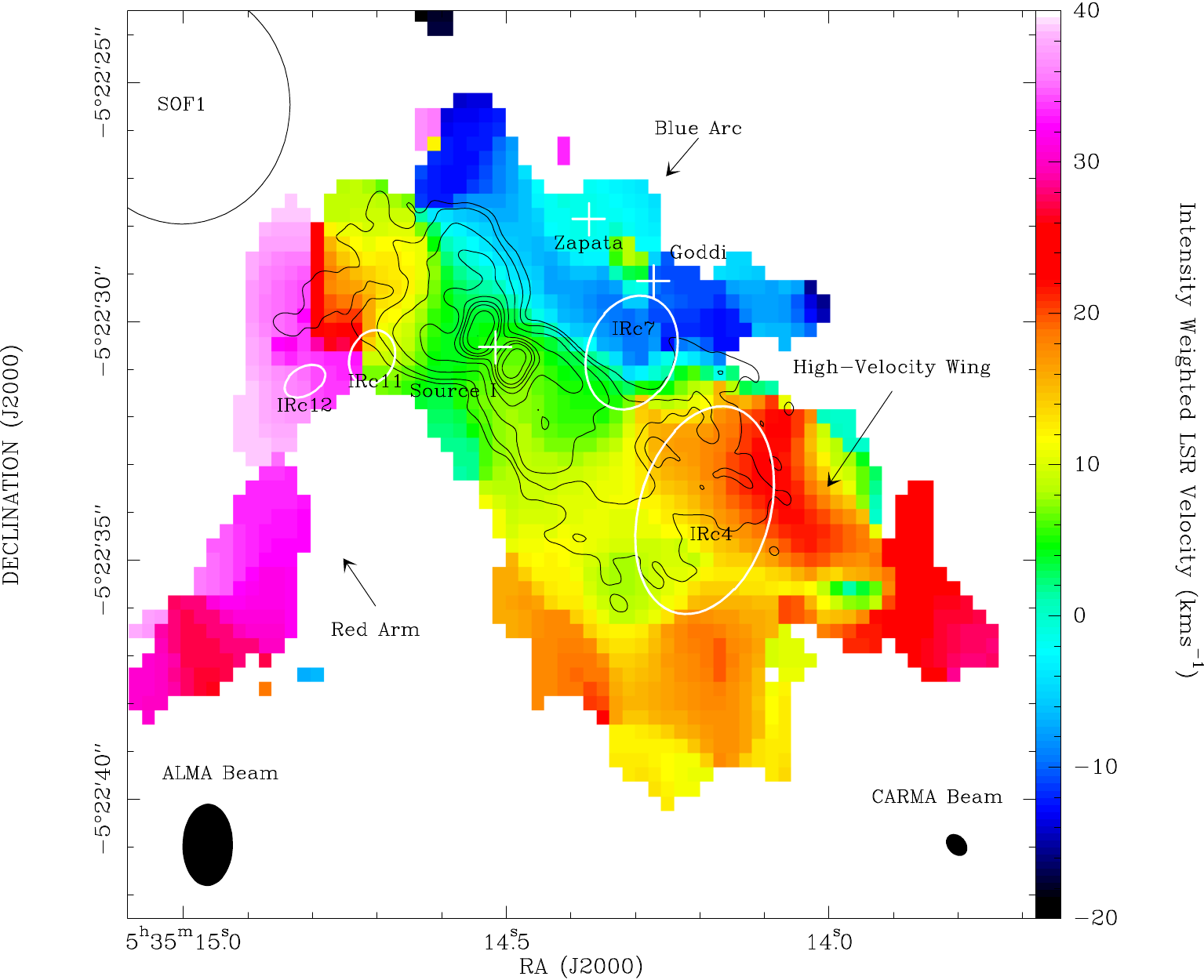}
\caption{Intensity-weighted velocity (moment 1) map of the v=0 J=5-4 $^{28}$SiO line emission with overlaid contours from the v=0 J=2-1 $^{28}$SiO total intensity \citep{Plambeck09}. Only pixels with fluxes $>$1 Jy~beam$^{-1}$ or 4 times the rms noise of the channel with the brightest emission, were used. The white ellipses indicate the positions \citep{Shuping04} and approximate sizes (estimated from \citealt{Shuping04}) of the infrared sources IRc4, IRc7, IRc11 and IRc12. 
 Source I is marked with a white cross. The newly discovered infrared source SOF1 \citep{deBuizer12} is indicated as a circle in the upper left corner. The two white crosses in the blue arc labeled `Goddi' and `Zapata' are the points of interaction between Source~I and BN as derived by \citet{Goddi11a} and \citet{Zapata09}, respectively. \label{28SiOv=0mom1}}
\end{figure}

The $^{28}$SiO v=0 J=5-4 line is the strongest and most extended line of all. In contrast to the other SiO lines, it does not have a pronounced double-peaked shape (Figure \ref{ALMA_spectra}). The peak of the emission is at about 6 km~s$^{-1}$ which is, with the given spectral resolution of 0.7 km~s$^{-1}$, consistent with the systemic velocity of Source I ($\sim$5.5 km~s$^{-1}$). The spectral profile shows an additional shoulder at 17 km~s$^{-1}$. The whole line emission is quite broad in velocity with a FWHM of about 22 km~s$^{-1}$. The channel maps for  this line emission are presented in Figure \ref{28SiOv=0_data_cube}. The $^{28}$SiO v=0 J=5-4 line emission traces an elongated outflow structure in the northeast-southwest direction and peaks at the position of Source~I for velocities smaller than 20 km~s$^{-1}$. At velocities $\lesssim$ 10 km~s$^{-1}$ the channel maps show an arc-like structure north of the position of Source~I. In addition, there is also emission at velocities $>$ 20 km~s$^{-1}$ that is not associated with Source~I. The position of the bulk of this high-velocity emission considerably overlaps with the mid-infrared source IRc4. For analysis purposes, we created maps of the velocity-integrated emission (moment 0; Figure \ref{28SiOv=0mom0}) and the intensity-weighted velocity (moment 1; Figure \ref{28SiOv=0mom1}). The moment 0 map is obtained integrating the emission at each pixel in the velocity range from -20 to 40 km~s$^{-1}$, while the moment 1 shows only pixels containing emission that is above a threshold of 4 $\sigma$ (= 1.08 Jy~beam$^{-1}$).

\subsection{The $^{29}$SiO v=0 Line}

\begin{figure}
\centering
\includegraphics[width=9cm]{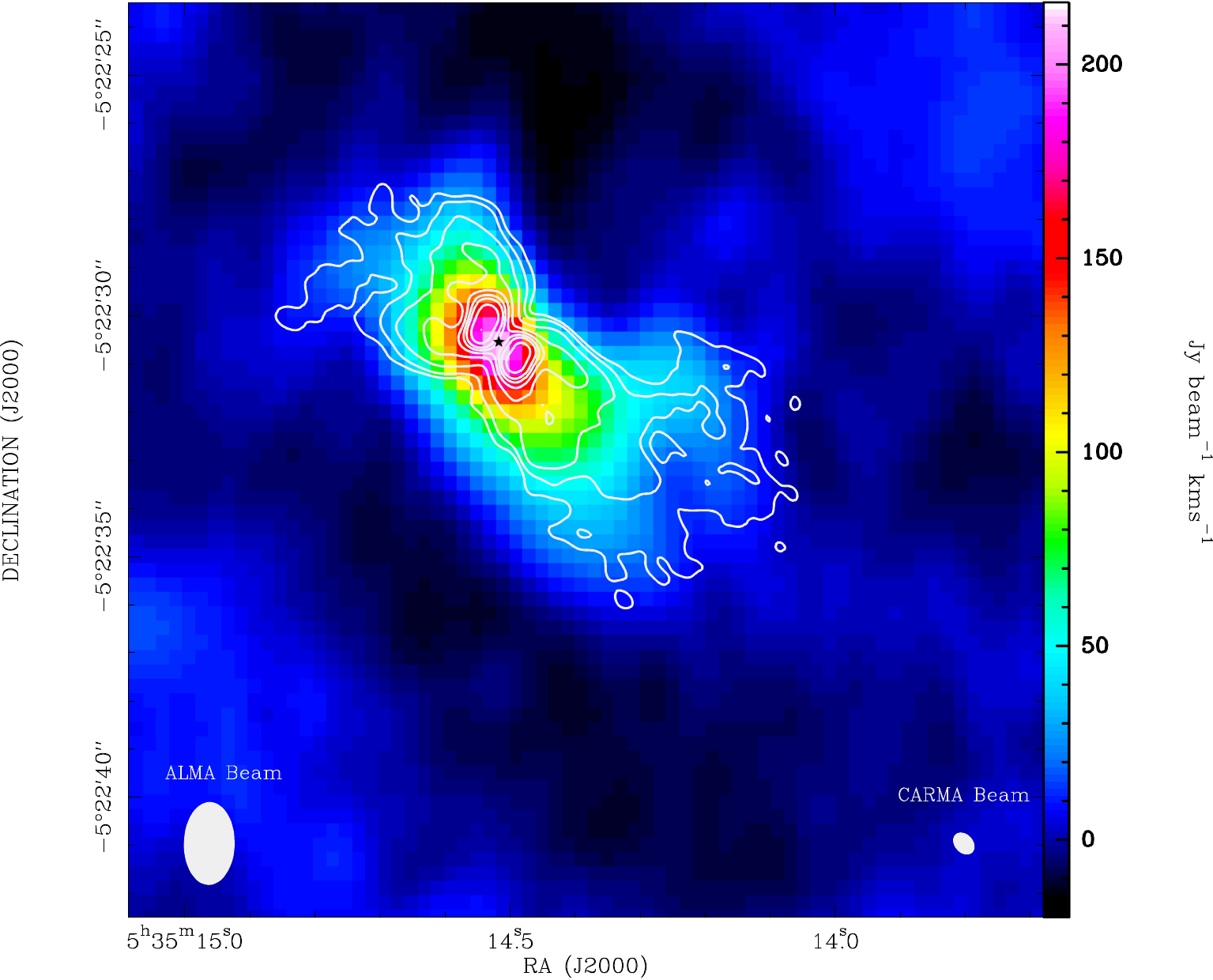}
\caption{Moment 0 map of the v=0 J=5-4 $^{29}$SiO transition overplotted with contours of the v=0 J=2-1 $^{28}$SiO total intensity. The contour levels are the same as in Figure \ref{28SiOv=0mom0}. The star symbol indicates the position of Source~I. \label{29SiOv=0mom0}}
\end{figure}

\begin{figure}
\centering
\includegraphics[width=9cm]{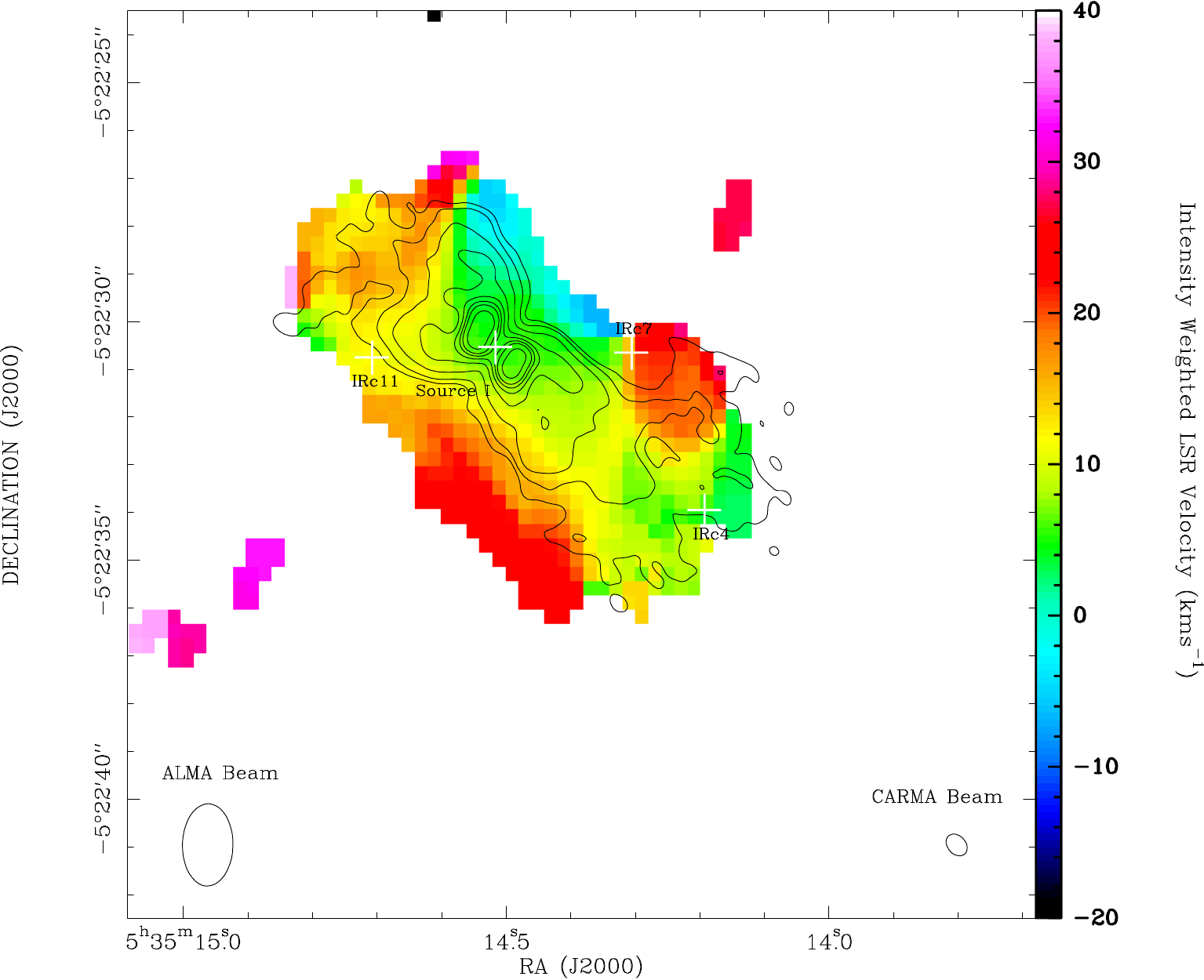}
\caption{Moment 1 map of the v=0 J=5-4 $^{29}$SiO line with overlaid contours from the v=0 J=2-1 $^{28}$SiO total intensity. Only pixels with fluxes larger than 4 times the rms noise of 0.17 Jy~beam$^{-1}$ were used to produce this map. The crosses mark the positions of IRc11, Source~I, IRc7 and IRc4. \label{29SiOv=0mom1}}
\end{figure}

The spectrum of the $^{29}$SiO v=0 J=5-4 transition also shows a double-peaked profile with two broad peaks at $\sim$1 km~s$^{-1}$ and $\sim$14 km~s$^{-1}$ (Figure \ref{ALMA_spectra}). The emission shows a similar velocity extent to the $^{28}$SiO v=0 J=5-4 transition. The blue end of the $^{29}$SiO spectrum appears to be contaminated by emission from the methyl-cyanide isotopologue $^{13}$CH$_3$CN originating from the hot core south of Source I. The channel maps (Figure \ref{29SiOv=0_data_cube}) show an emission structure similar to the v=0 $^{28}$SiO J=5-4 emission. As for the $^{28}$SiO v=0 line, we created maps of the velocity-integrated emission and intensity-weighted velocity (Figures \ref{29SiOv=0mom0} and \ref{29SiOv=0mom1}). The moment 0 image shows the velocity-integrated emission from every pixel and the moment 1 map contains only pixels above 4 $\sigma$ (= 0.68 Jy~beam$^{-1}$).  

\begin{figure}
\centering
\includegraphics[width=9cm]{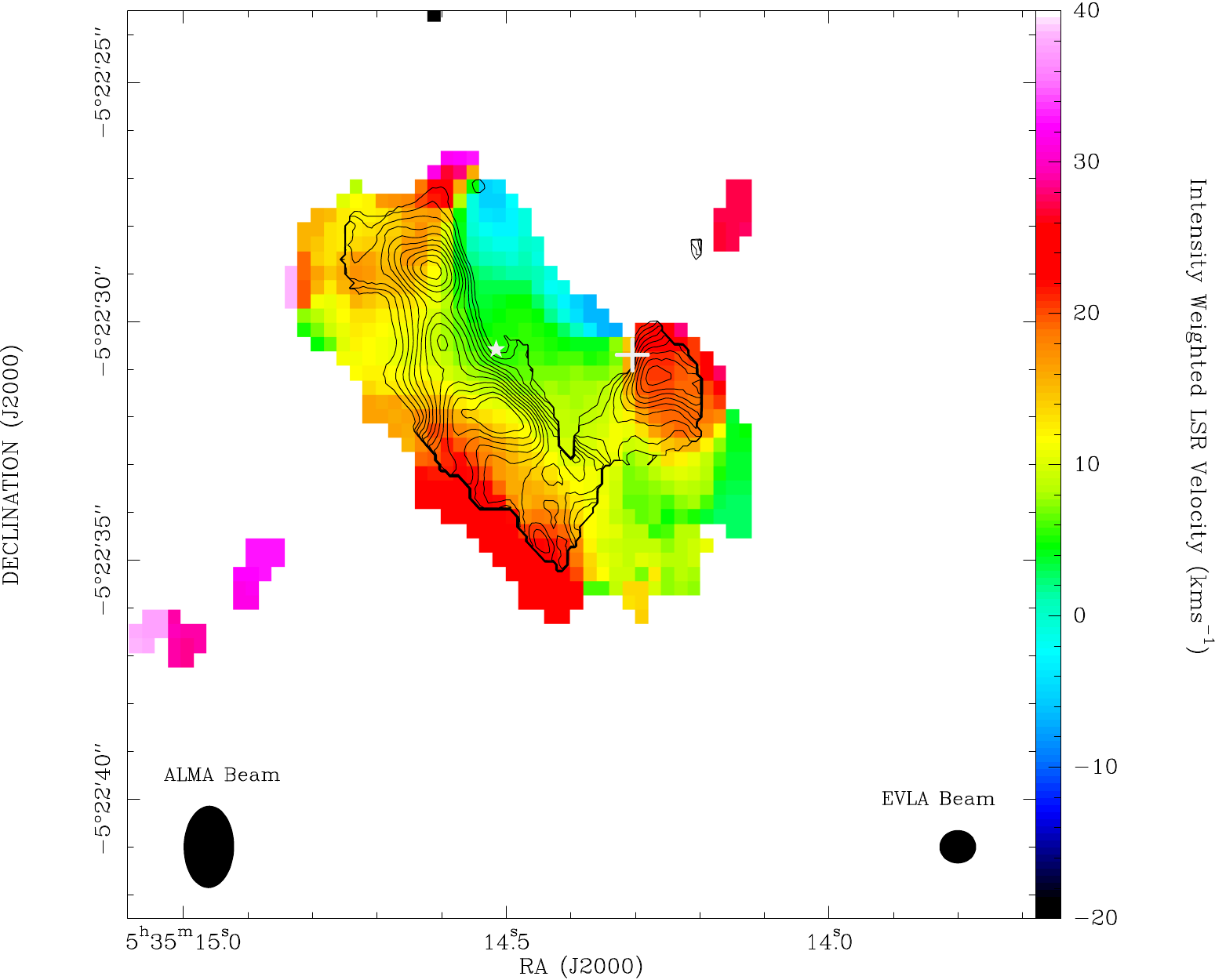}
\caption{Moment 1 map of the v=0 J=5-4 $^{29}$SiO line overlaid with contours of the ammonia column density map from EVLA observations \citep{Goddi11b}. The contours are from 35\% to 90\% of the peak column density of 2.0$\times$10$^{17}$ cm$^{-2}$, in steps of 5\%. The star symbol indicates the position of Source~I and the cross marks the position of IRc7. \label{ammonia_overlay}}
\end{figure}

\subsection{The $^{28}$Si$^{18}$O v=0 Line}

The $^{28}$Si$^{18}$O spectral profile shows a double-peaked line shape similar to the other isotopologues. The blue-shifted peak, however, appears to be highly blended by CH$_3$CH$_2$CN emission originating from the Hot Core (see for example the channel maps in Figure \ref{28Si18Ov=0_data_cube}). We cannot determine the exact velocity range of the SiO emission, because of the strong blending in the blue part of the emission. Nevertheless, from the channel maps one can estimate that the SiO emission ranges from about 2 km~s$^{-1}$ up to 17 km~s$^{-1}$. In this velocity range, the $^{28}$Si$^{18}$O emission shows a structure that is similar to, but weaker than the $^{29}$SiO line emission.

\section{Discussion\label{sec:disc}}

\subsection{The Structure and Velocity Field of the Outflow}

The outflow from Source~I has a well-ordered spatial and velocity structure within 1000 AU, as evidenced by high-angular resolution imaging of the SiO J=1-0 rotational lines in the vibrationally-excited states \citep{Matthews10} and the ground-state \citep{Greenhill12}. In particular, the flow appears as a wide-angle rotating wind within 100 AU from Source~I, and then collimates into a bipolar, linear, rotating flow at radii of 100-1000 AU. The new ALMA images of the J=5-4 rotational lines from different isotopologues reported here, probe larger scales of the flow, up to $\sim$5000 AU. 
Interestingly, these large scales show a rather complex outflow morphology and velocity field when compared with the innermost flow regions (Figures \ref{28SiOv=0mom0}, \ref{28SiOv=0mom1}, \ref{29SiOv=0mom0}, \ref{29SiOv=0mom1}). 
We interpret this peculiar behaviour on different scales as a result of the interaction of the outflow from Source~I with the complex environment of BN/KL. In the following subsections, we will focus on investigating the relation of  the ground-state $^{28}$SiO and $^{29}$SiO emission probing the large-scales of the outflow with the infrared sources located in the vicinity of Source~I, the explosive event occurred about 500 years in BN/KL, and dense gas in the Orion hot core.

\subsubsection{The v=0 $^{28}$SiO Emission}

Figures \ref{28SiOv=0mom0} and \ref{28SiOv=0mom1} show the moment 0 and moment 1 maps of the $^{28}$SiO J=5-4 transition observed with ALMA, overlaid with the contours of the total intensity of the $^{28}$SiO v=0 J=2-1 emission as observed by \citet{Plambeck09} using CARMA. The 5-4 and 2-1 transitions of $^{28}$SiO are broadly consistent both in morphology and position angle, and clearly trace a bipolar outflow driven by Source~I into its surrounding molecular cloud along a northeast-southwest axis. 
Although the outflow has a pronounced bow-tie morphology, it is not symmetric. The southern lobe appears broader and more extended than the northern one, whereas, perpendicularly to the outflow axis, the emission in both lobes is more extended to the north-west (Figure \ref{28SiOv=0mom0}). 

The velocity field in the moment 1 map displays more interesting details on the large-scale structure of the outflow (Figure \ref{28SiOv=0mom1}). 
In particular, three new elements are observed. 

\begin{enumerate}[label=\roman*)]

\item
 The southern lobe displays two wings of red-shifted emission, the first one to the east is seen in the velocity range from $\sim$7.5 km~s$^{-1}$ to $\sim$17.5 km~s$^{-1}$ and the second one to the west traces the highest velocity red-shifted gas, from $\sim$20 km~s$^{-1}$ to $\sim$35 km~s$^{-1}$ (``high-velocity wing"; Figure \ref{28SiOv=0mom1}). 

\item
 The end of the north  lobe of the outflow appears to be ``bent'' to the east resulting in an arm-like structure of faint highly red-shifted ($\gtrsim$25 km~s$^{-1}$) emission, having a northwest-southeast orientation  (``red arm"; Figure \ref{28SiOv=0mom1}). 

\item
 An arc-like feature of blue-shifted emission ($\sim$-10 km~s$^{-1}$ to $\sim$5 km~s$^{-1}$), centered on Source~I, extends along the northwest-facing edge of the outflow (``blue arc"; Figure \ref{28SiOv=0mom1}). 

\end{enumerate}

Since these features are evident in other thermal transitions of SiO, like the v=0 J=2-1 \citep{Plambeck09} and the v=0 J=8-7 \citep{Zapata12}, we exclude contamination from other molecular species, i.e. they must be associated with SiO emission.

Consistently with our findings, \citet{Zapata12} noticed that the J=5-4 v=0 SiO emission has a butterfly-like shape with the blue-shifted emission in the northwest-facing side of the outflow (corresponding to our blue arc) and the red-shifted emission in the southeast-facing side (corresponding to our moderately high-velocity wing); they also noticed a  high-velocity ``tail" lying in between the two wings (see their Figure 3), which corresponds to our ``high-velocity wing" (Figure \ref{28SiOv=0mom1}). 
They discuss some possibilities to explain this peculiar velocity structure, in particular the simultaneous presence of two bipolar outflows from a binary or a single rotating  outflow. Two outflows having north-south and east-west axes, respectively, could explain the blue and red wings but can be ruled out by the presence of the highest red-shifted tail or wing in the southern lobe. The second option of a single outflow assumes a sense of rotation   with redshifted emission in the southeast-facing side and blue-shifted emission in the northwest-facing side of the outflow. On scales smaller than 100 AU, however, \citet{Matthews10} observed rotation in the opposite direction, requiring a change of the rotation sense from small to large scales, which is unlikely. 
Precession in a binary over a timescales of several hundreds of years could in principle explain the difference in the red-shifted and blue-shifted  velocities at different scales \citep{Plambeck09}, but it is inconsistent with the strong evidence that Source I is a hard binary with an orbital separation of (at most) a few AU and an orbital period of a few years \citep{Goddi11a,Bally11}. 
Besides, while we see a velocity offset from red to blue in the northern lobe in the southeast-northwest direction, in the southern lobe we observe the opposite trend, with the more redshifted emission in the north-west (``high-velocity wing"; Figure \ref{28SiOv=0mom1}). 
The outflow rotation scenario faces a similar challenge in explaining the tail with the highest red-shifted emission lying in between the moderately red-shifted and the blue-shifted wings (\citealt{Zapata12}; their Figure 3).  
We conclude that on scales of thousands of AU there is not evidence for a clear velocity gradient perpendicular to the outflow axis that can be interpreted as rotation. 

\begin{figure}
\centering
\includegraphics[width=7cm]{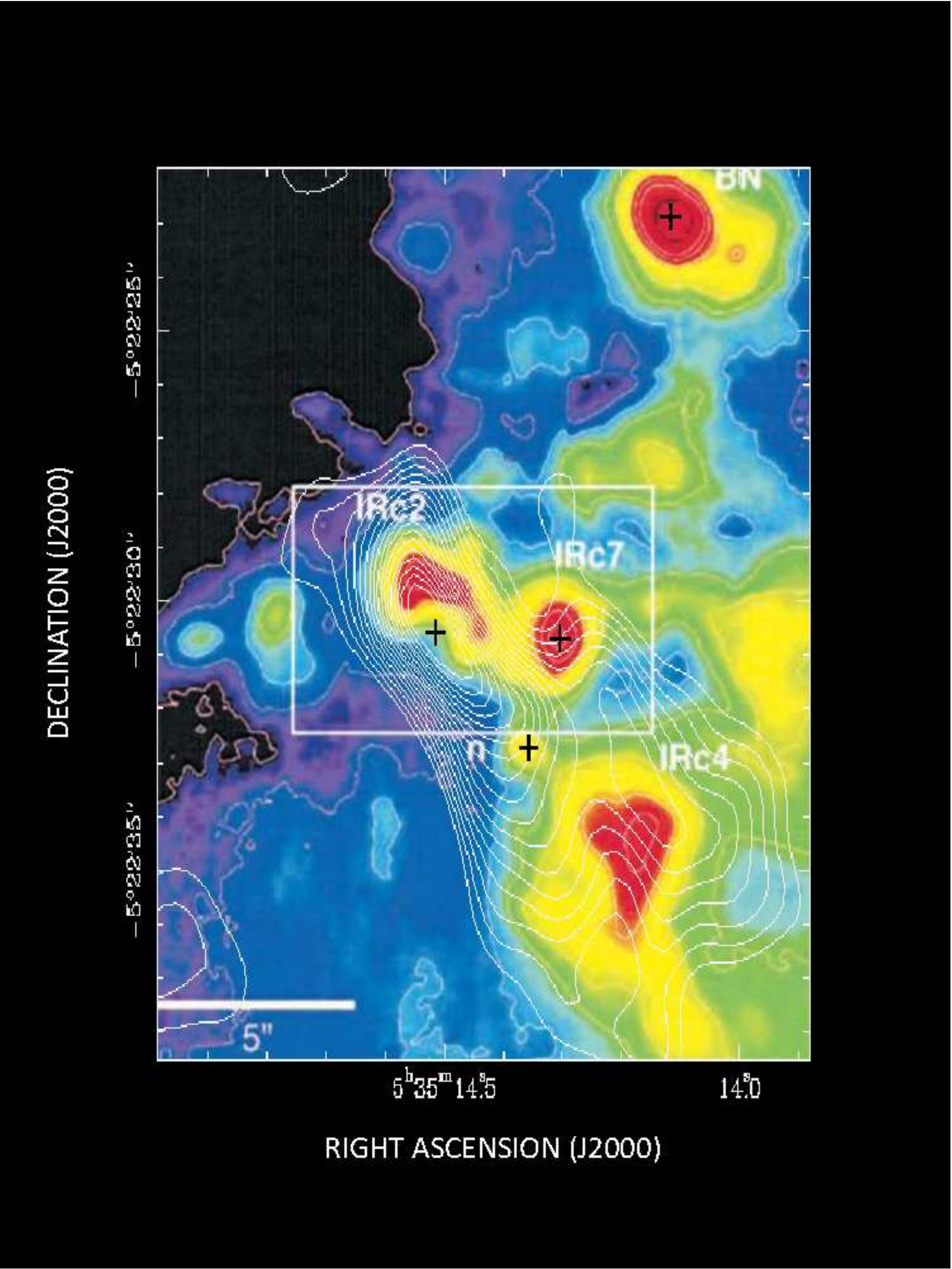}
\caption{Orion BN/KL 12.5 $\mu$m mosaic \citep{Greenhill04a} overlaid with v=0 J=5-4 $^{28}$SiO integrated intensity contours. The contours are from 10\% to 90\% of the peak emission of 170 Jy~beam$^{-1}$, in steps of 5\%. The crosses indicate the positions of Source~I, Source~n, IRc7 and the BN object (from left to right). \label{gemini_overlay}}
\end{figure}

We propose an alternative explanation, that 
the peculiar features observed in the SiO velocity field on large scales are most likely related to the outflow interaction with its environment. 
For this purpose, we overlaid the SiO velocity field  map with mid-infrared sources in the vicinity of Source I (Figures \ref{28SiOv=0mom1} and \ref{gemini_overlay}). 
These overlays evidence three interesting elements. 

\begin{enumerate}[label=\roman*)]

\item
  The contours of the southern lobe of the SiO emission clearly follow the 12 $\mu$m emission from IRc4 (Figure \ref{gemini_overlay}) and the high-velocity red wing significantly overlaps with IRc4 (Figure \ref{28SiOv=0mom1}). Using SOFIA, \citet{deBuizer12} propose from their IR observations of the BN/KL region that IRc4 is self-luminous. This high red-shifted wing of SiO emission may not be associated with the outflow of Source~I but with IRc4, that harbors a YSO that could heat the dust in its vicinity and enrich SiO via sublimation. Alternatively, the interaction of the Source~I outflow with IRc4 could create shocks that sputter dust grains enriching gas phase SiO. 

\item
The northern part of the bended high-velocity ``red arm" arising from the north lobe to the east of Source~I overlaps well 
with the positions of the mid-infrared sources IRc11 and IRc12, which may play a role in exciting this high-velocity SiO emission. IRc11 and IRc12 alone, however, cannot explain the whole ``red arm". Since the latter has a filamentary structure with a northwest-southeast orientation, reminiscent of the ``fingers" of shocked H$_2$ bullets in the fast OMC-1 outflow, we speculate that the ``red arm" (in particular its southernmost part) may be tracing part of the high-velocity northwest-southeast outflow that originated from a close encounter of Source~I and the BN object approximately 500 years ago \citep{Goddi11a,Bally11}. 

\item
The center of the ``blue arc" 
to the north-west of Source~I falls in between two independent measurements of the center of dynamical interaction between Source~I and the BN object (\citealt{Goddi11a,Zapata09}; Figure \ref{28SiOv=0mom1}). We speculate that this blue-shifted emission, similarly to the red-shifted emission from the ``red arm", could arise from remnant material that was stripped away from the interacting objects with moderately velocity.   

\end{enumerate}

\subsubsection{The v=0 $^{29}$SiO Emission}
Figures \ref{29SiOv=0mom0} and \ref{29SiOv=0mom1} show the ALMA moment 0 and moment 1 maps of the $^{29}$SiO J=5-4 transition, overlaid with the contours of the total intensity of the CARMA $^{28}$SiO v=0 J=2-1 emission \citep{Plambeck09}.  The structure of $^{29}$SiO  J=5-4 line follows the $^{28}$SiO J=2-1 contours by \citet{Plambeck09} quite well (Figure \ref{29SiOv=0mom0}).  The $^{29}$SiO  J=5-4 emission appears more compact than the $^{28}$SiO  J=5-4 emission but has an overall similar morphology, with the southern lobe of the outflow being more extended than the northern lobe (e.g. compare Figures \ref{28SiOv=0mom0} and \ref{29SiOv=0mom0}). 

The velocity field shows a much clearer blue-red asymmetry than the $^{28}$SiO   perpendicular to the outflow axis (along northwest-southeast). This apparent velocity gradient could be in principle interpreted as rotation. 
We argue, however, that this asymmetry, as well as the one displayed in the $^{28}$SiO emission, cannot be attributed to rotation but most probably to the interaction of the outflow with its environment.
In particular, the blue-shifted emission along the northwest-facing edge of the outflow is observed also in the $^{28}$SiO emission, for which we favor an association with the explosive event rather than Source~I itself. 
On the contrary, the red-shifted emission (with velocities $>$ 20 km~s$^{-1}$)  is mainly composed of two red features that are not seen in the $^{28}$SiO J=5-4 line or in the J=2-1 line by \citet{Plambeck09}: 1) an elongated structure in the southeast-facing edge of the outflow; 2) a red clump near the position of IRc7. As this emission is not seen in the other SiO transitions, it may be associated with another contaminant molecule rather than SiO. 
Figure \ref{ammonia_overlay} shows 
the moment 1 map of the $^{29}$SiO J=5-4 transition overlaid with contours of the ammonia column density map \citep{Goddi11b} tracing dense gas in the hot core region. The contours appear to overlap significantly with the two red features.

Since the high red-shifted $^{29}$SiO emission seems to follow the dense gas traced by the ammonia column density
 (Figure \ref{ammonia_overlay}), we suggest that it may trace the interface of the Source~I outflow with the hot core and IRc7, respectively. 

Consistently, \citet{Tercero11} showed that the $^{29}$SiO v=0 J=5-4 line is contaminated by the following molecules: the hydrogen-sulfide isotopologue $^{34}$SH$_2$, the methyl-cyanide isotopologue $^{13}$CH$_3$CN, and sulfur-monoxide (SO). \citet{Beuther05} imaged the (19$_{6,19}$-18$_{6,18}$) $^{13}$CH$_3$CN line at 339,137 GHz, which clearly peaks both at the hot core and  IRc7 positions. If the $^{13}$CH$_3$CN emission is responsible for the strong transition at 214.37437 GHz, we estimate an LSR velocity of $\gtrsim$ 6 km~s$^{-1}$ which is reasonable for emission coming from the hot core (v$\sim$9 km~s$^{-1}$). We therefore propose that the apparent high-velocity features in the $^{29}$SiO maps can be attributed  to $^{13}$CH$_3$CN and be associated with IRc7 and the interface between the hot core and the outflow from Source I. 

We conclude that the blue-red asymmetry in the SiO moment 1 map is only apparent. Therefore, we exclude the presence of a velocity gradient perpendicular to the outflow axis as well as evidence of outflow rotation on scales of thousands of AU from Source~I. 

\subsection{The Nature of the Emission \label{sec:nature}}
In order to investigate the nature of the emission (thermal vs. maser), we estimated the brightness temperature $ T_b$ of the $^{28}$SiO v=0, v=1 and the $^{29}$SiO v=0 transitions (Table \ref{ALMA_Data}). 
$T_b$ is given by:
$T_b=(S\lambda^2)/(2k_Bd\Omega)$,
where $S$ is the flux, $\lambda$ the wavelength and $d\Omega$ the solid angle of the source.
The $^{28}$SiO v=1 emission is not resolved, therefore we can only give a lower limit to the brightness temperature. For a beamsize of 1$\farcs$7$\times$1$\farcs$2 and a peak flux of 45 Jy, we derive a brightness temperature $T_b\sim$580 K. 
Based on the lower limit of the brightness temperature alone, the nature of the v=1 emission remains ambiguous.
However, the two spectral peaks of the $^{28}$SiO v=1 transition are narrow (line-widths 3.5 km~s$^{-1}$, Figure \ref{ALMA_spectra}), consistent with maser emission. Moreover, the positions fitted to the unresolved emission at different velocities show a distribution similar to the J=1-0 v=1 masing transition \citep{Matthews10} (see Section 3.1 and Figure \ref{spot_pos}). These elements provide first evidence that the v=1 J=5-4 $^{28}$SiO transition is a maser. 
If confirmed, this would be the first detection of high-frequency SiO maser emission ($>$ 200 GHz) associated with Source I. In addition, the masing nature of the v=1 J=5-4 transition makes it a valuable tracer of the inner ($<$ 100 AU) part of the Source~I outflow for follow up observations using ALMA long baselines (15 km).

The v=0 lines from $^{28}$SiO and $^{29}$SiO are resolved, so we calculated $T_b$ by integrating the flux over the angular size of the source. We obtained brightness temperatures of $\sim$50 K and $\sim$100 K for the $^{28}$SiO and the $^{29}$SiO isotopologue, respectively. 
The low brightness temperature of the v=0 $^{28}$SiO and $^{29}$SiO transitions indicates that the emission is most likely thermal. 
But the fact that the emission from the $^{29}$SiO isotopologue is brighter than the one arising from the $^{28}$SiO molecule, suggests that the $^{29}$SiO transition could be partly masing.

\section{Conclusions and Future Work\label{sec:concl}}

We have reported ALMA observations of five rotational transitions of three SiO isotopologues towards the Orion BN/KL region. The vibrational ground-state emission from $^{28}$SiO and $^{29}$SiO shows an extended bipolar structure along a northeast-southwest axis at the position of Source~I and traces the outer parts of its outflow up to scales of a few thousand AU. These large-scales reveal complex morphological and dynamical structures that may be due to the outflow interaction with its environment, in particular the hot core and the mid-infrared sources in the vicinity.
Emission from the vibrationally-excited v=1 and v=2 J=5-4 $^{28}$SiO is not resolved with the current resolution ($\sim$1$\farcs$5; $\sim$600 AU at 420 pc). 2-D Gaussian fitting to the unresolved emission in each velocity channel indicates it traces the inner regions ($<$100 AU) of the outflow associated with Source~I.
Based on the narrow line-widths of spectral features ($<$3.5 km~s$^{-1}$) 
and a distribution similar to the masing $^{28}$SiO J=1-0 v=1 line, we found first evidence for the $^{28}$SiO v=1 J=5-4 transition being masing. Follow-up observations with the full ALMA array will enable resolution of the region of the $^{28}$SiO v=1 J=5-4 emission and provide data complementary to those taken at lower frequencies, possibly unveiling new dynamical and physical properties of molecular gas in a disk/outflow system associated with the closest known high-mass YSO.

\begin{acknowledgements} 
This paper makes use of the following ALMA data: ADS/JAO.ALMA\#2011.0.99001.CSV.
ALMA is a partnership of ESO (representing its member states), NSF (USA) and NINS (Japan),
together with NRC (Canada) and NSC and ASIAA (Taiwan), in cooperation with the Republic of Chile.
The Joint ALMA Observatory is operated by ESO, AUI/NRAO and NAOJ. We thank the ALMA
Commissioning and Science Verification team, Department of Science Operations and ARC staff
for their work to make these data available. We also thank Roberto Galvan-Madrid for useful discussions.
\end{acknowledgements}

\end{document}